\documentclass[10pt,journal,compsoc]{IEEEtran}
\usepackage{booktabs} 
\usepackage{amsmath}
\usepackage{graphicx}
\usepackage[belowskip=-12pt]{caption}
\usepackage{rotating}
\usepackage{tabularx}
\usepackage{subfig}
\usepackage{wrapfig}
\usepackage{listings}
\usepackage{balance} 
\usepackage{lscape}
\usepackage[table]{xcolor}
\usepackage{hyperref}
\usepackage{xspace}
\usepackage{framed}
\usepackage{syntax}
\usepackage{hyperref}
\usepackage[framemethod=TikZ]{mdframed}

\usepackage{flushend}

\definecolor{codegreen}{rgb}{0,0.6,0}
\definecolor{codegray}{rgb}{0.5,0.5,0.5}
\definecolor{codepurple}{rgb}{0.58,0,0.82}
\definecolor{backcolour}{rgb}{0.95,0.95,0.92}
\definecolor{Gray}{gray}{0.1}
\lstdefinestyle{mystyle}{
	backgroundcolor=\color{backcolour},   
	commentstyle=\color{codegreen},
	keywordstyle=\color{magenta},
	numberstyle=\tiny\color{codegray},
	stringstyle=\color{codepurple},
	basicstyle=\scriptsize,
	breakatwhitespace=false,         
	breaklines=true,                 
	captionpos=b,                    
	keepspaces=true,                 
	numbers=left,                    
	numbersep=5pt,                  
	showspaces=false,                
	showstringspaces=false,
	showtabs=false,                  
	tabsize=2
}

\newcommand{\change}[1]{{\color{blue!80!white}\textbf{}~#1}\xspace}

\graphicspath{{./figures/}}
\newcommand{\figref}[1]{Fig.~\ref{#1}}
\newcommand{\tabref}[1]{Table~\ref{#1}}
\newcommand{\fignref}[1]{Figure~\ref{#1}}

\newcommand{\secref}[1]{Section~\ref{#1}}

\newcommand{\etal}{{\em et al.}\xspace}

\newcommand{\code}[1]{{\texttt{\footnotesize{#1}}\xspace}}

\newcommand{\sof}{\textit{Stack Overflow}\xspace}
\newcommand{\caffe}{\textit{Caffe}\xspace}
\newcommand{\ho}{\textit{H2O}\xspace}
\newcommand{\keras}{\textit{Keras}\xspace}
\newcommand{\mahout}{\textit{Mahout}\xspace}
\newcommand{\mllib}{\textit{MLlib}\xspace}
\newcommand{\scikit}{\textit{scikit-learn}\xspace}
\newcommand{\tensor}{\textit{Tensorflow}\xspace}
\newcommand{\theano}{\textit{Theano}\xspace}
\newcommand{\torch}{\textit{Torch}\xspace}
\newcommand{\weka}{\textit{Weka}\xspace}

\newcounter{NumObservations}
\stepcounter{NumObservations}

\newcommand{\findings}[2]{%
	\begin{mdframed}[backgroundcolor=gray!10,
		linewidth=1pt,
		roundcorner=5pt]%
		{\bf{Finding \arabic{NumObservations}}}: #1 \\
		\vspace{-0.85em}%
	\end{mdframed}%
	\stepcounter{NumObservations}
}

\ifCLASSOPTIONcompsoc
\usepackage[nocompress]{cite}
\else
\usepackage{cite}
\fi

\ifCLASSINFOpdf

\else

\fi

\begin{document}
	\title{What Do Developers Ask About ML Libraries? A Large-scale Study Using Stack Overflow}
	
	
	\author{Md~Johirul~Islam,~
		Hoan~Anh~Nguyen,~
		Rangeet~Pan,
		Hridesh~Rajan ~
		\IEEEcompsocitemizethanks{\IEEEcompsocthanksitem Md Johirul Islam is with the Department
			of Computer Science, Iowa State University, Ames,
			IA, 50010.\protect\\
			E-mail: mislam@iastate.edu
		}
		\IEEEcompsocitemizethanks{\IEEEcompsocthanksitem Hoan Anh Nguyen is with the Department
			of Computer Science, Iowa State University, Ames,
			IA, 50010.\protect\\
			E-mail: hoan@iastate.edu
		}
	
	\IEEEcompsocitemizethanks{\IEEEcompsocthanksitem Rangeet Pan is with the Department
		of Computer Science, Iowa State University, Ames,
		IA, 50010.\protect\\
		E-mail: rangeet@iastate.edu
	}

		\IEEEcompsocitemizethanks{\IEEEcompsocthanksitem Hridesh Rajan is with the Department
			of Computer Science, Iowa State University, Ames,
			IA, 50010.\protect\\
			E-mail: hridesh@iastate.edu
		}
	}

	\IEEEtitleabstractindextext{%
		\begin{abstract}

Modern software systems are increasingly including machine learning (ML) as 
an integral component. However, we do not yet understand the difficulties 
faced by software developers when learning about ML libraries and using them 
within their systems. To that end, this work reports on a detailed (manual) 
examination of 3,243 highly-rated Q\&A posts related to ten ML libraries, 
namely \tensor, \keras, \scikit, \weka, \caffe, \theano, \mllib, \torch, 
\mahout, and \ho, on \sof, a popular online technical Q\&A forum. We classify 
these questions into seven typical stages of an ML pipeline to understand the 
correlation between the library and the stage.
Then we study the questions and perform statistical analysis to explore the answer to four research objectives (finding the most difficult stage, understanding the nature of problems, nature of libraries and studying whether the difficulties stayed consistent over time).
 Our findings reveal the urgent 
need for software engineering (SE) research in this area. Both static and dynamic 
analyses are mostly absent and badly needed to help developers find errors 
earlier. While there has been some early research on debugging, much more work is 
needed. API misuses are prevalent and API design improvements are sorely 
needed.  
Last and somewhat surprisingly, a tug of war between providing higher levels 
of abstractions and the need to understand the behavior of the trained model 
is prevalent. 

\end{abstract}

		}

	\maketitle

	\IEEEdisplaynontitleabstractindextext
	
	\IEEEpeerreviewmaketitle

	\section{Introduction}\label{sec:intro}

Machine learning (ML) is becoming an essential computational tool 
in a software developer's toolbox for solving problems that defy traditional 
algorithmic approach.
Software developers are fulfilling this need by development and refinement 
of a number of new ML libraries~\cite{ml16}.
Recently it has also been suggested that ML can introduce unique software
development problems~\cite{sculley2014machine,sculley2015hidden,breck2016s}.
However, we do not yet know about the problems that users of ML libraries
face and those that they choose to ask about publicly.

Prior work has shown that studying question and answer (Q\&A) forums such 
as \sof can give significant insights into software developer's 
concerns about a technology~\cite{treude2011programmers,%
	wang2013empirical,yang2014sparrows,kavaler2013using,linares2014api,sahu2016selecting,%
	barua2014developers,wang2013detecting,rebouccas2016empirical,zhang2015evaluating,%
	schenk2013geo,stanley2013predicting,mcdonnell2013empirical,baltadzhieva2015predicting,%
	joorabchi2016text}, but has not focused on ML libraries. 
More details of related work are discussed in \secref{sec:related}. 

This work presents a study of the problems faced by developers while using 
popular ML libraries. Our study also leverages the posts on \sof. Since 2015, 
there has been growing interest and significant increase in ML related 
questions and distinct users making \sof a representative source of dataset 
for our study. We selected 10 ML libraries to study, identified by a 
survey~\cite{ml16} and confirmed by counting the number of posts on \sof related to 
those libraries.
These libraries 
are 
\caffe~\cite{jia2014caffe},
\ho~\cite{candel2016deep},
\keras~\cite{chollet2015keras},
\mahout~\cite{owen2012mahout}, 
\mllib~\cite{meng2016mllib},
\scikit~\cite{pedregosa2011scikit},
\tensor~\cite{abadi2016tensorflow},
\theano~\cite{bergstra2011theano},
\torch~\cite{collobert2002torch},
and \weka~\cite{holmes1994weka}.

\textbf{\caffe}~\cite{jia2014caffe} is a deep learning library for 
Python and C++. 
\textbf{\ho}~\cite{candel2016deep} is a deep learning library for Java, R, Python
or Scala and its key feature is to provide a workflow-like system for building ML
models. 
\textbf{\keras}~\cite{chollet2015keras} is a deep learning library for Python
whose key feature is to provide higher-level abstractions to make
creating neural networks easier. \keras also uses \tensor or \theano as the backend.
\textbf{\mahout}~\cite{owen2012mahout} is aimed at 
providing scalable ML facilities for Hadoop clusters. 
\textbf{\mllib}~\cite{meng2016mllib} is aimed at providing
scalable ML facilities for Spark clusters. 
\textbf{\scikit}~\cite{pedregosa2011scikit} is a Python library 
that uses \tensor or \theano as the backend. This library provides a rich set
of abstract APIs~\cite{sklearnapi} to hide complexity of ML from
the user in an effort to make ML features widely accessible.
\textbf{\tensor}~\cite{abadi2016tensorflow} provides facilities to 
represent a ML model as data flow graphs.
\textbf{\theano}~\cite{bergstra2011theano} and \textbf{\torch}~\cite{collobert2002torch}
are aimed at scaling ML algorithms using GPU computing.
A novelty of \theano is that it provides some self-verification and
unit testing to diagnose some runtime errors.
\textbf{\weka}~\cite{holmes1994weka}
is a ML library for Java. It provides API
support for data preparation, classification, regression, clustering and
association rules mining tasks and a GUI for making models easier. 

All in all, this set is both representative and provides variety.
We selected a total of 3,243 highly-rated \sof posts for this study. A team 
of three Ph.D. students, with experience in coursework on AI and ML, and 
using ML libraries, independently read and labeled each of the posts 
producing 9,849 labels that were then compared for consistency producing 177 
conflicting labels on 177 different posts.
All of these conflicts were resolved using mediated, face-to-face 
conflict resolution meetings between all three participants.
We then performed a statistical analysis and a study of the data to answer the following research questions: \newline
\textbf{RQ1: Difficult stage } Which stages are more difficult in a ML pipeline? 
\figref{fig:steps} shows stages in a typical ML pipeline. \newline
\textbf{RQ2: Nature of problems} Which  problems are more specific to library and which are inherent to ML? \newline
\textbf{RQ3: Nature of libraries} Which libraries face problems in specific stages and which ones face difficulties in all stages?\newline
\textbf{RQ4: Consistency} Did the problems stay consistent over the time?

The remainder of this work describes our study and results and
makes the following contributions: 
(1) a labeled and verified, dataset of 3,243 ML library-related Q\&A on \sof,
(2) a classification scheme for ML-related Q\&A,
(3) an intra-library analysis to identify strengths and weaknesses of ML libraries, and
(4) an inter-library analysis to identify relative strengths and weaknesses. 

	\section{Methodology}
\label{sec:approach}

\begin{table}[t]
	\centering
\caption{Numbers of posts having different score (S) about ML libraries. The bold column represents selected posts.}
\setlength{\tabcolsep}{4.8pt}
\begin{tabular}{ |l r r r r r r| } 
 \hline
 \rowcolor{gray}
 \textcolor{white}{\bf Library}  & \textcolor{white}{\bf S $\ge$ 0}  & \textcolor{white}{\bf S $\ge$ 1} & \textcolor{white}{\bf S $\ge$ 2} & \textcolor{white}{\bf S $\ge$ 3} & \textcolor{white}{\bf S $\ge$ 4} & \textcolor{white}{\bf S $\ge$ 5} \\ 
 \hline
 \hline
 
 \caffe \cite{jia2014caffe} & 2,339 & 1,320 & 620 & 318 & 192 & {\bf 132} \\ 
  \rowcolor{lightgray}
 \ho \cite{candel2016deep} & 771 & 452 & 167 & 73 & 34 & {\bf 17} \\ 
\keras \cite{chollet2015keras} & 5,708 & 3,323 & 1,751 & 953 & 568 & {\bf 367} \\ 

  \rowcolor{lightgray}
  \mahout \cite{owen2012mahout} & 1,186 & 610 & 293 & 160 & 103 & {\bf 48} \\ 
\mllib \cite{meng2016mllib} & 1,688 & 929 & 498 & 272 & 173 & {\bf 119} \\ 
 
  \rowcolor{lightgray}
  \scikit \cite{pedregosa2011scikit} & 9,246 & 5,302 & 2,898 & 1,759 & 1,188 & {\bf 856} \\ 
 
\tensor\cite{abadi2016tensorflow} & 21,115 & 10,109 & 4,962 & 2,769 & 1,827 & {\bf 1,334} \\ 
 
  \rowcolor{lightgray}
  \theano \cite{bergstra2011theano} & 2,332 & 1,341 & 711 & 421 & 265 & {\bf 192} \\ 
 
 \torch \cite{collobert2002torch} & 1,226 & 640 & 312 & 161 & 91 & {\bf 61} \\ 
  \rowcolor{lightgray}
  \weka \cite{holmes1994weka} & 2,512 & 1,216 & 568 & 293 & 181 & {\bf 117} \\ 
 Total & 48,123 & 25,242 & 12,780 & 7,179 & 4,622 & {\bf 3,243} \\ 
 \hline
\end{tabular}
 \label{tbl:dataset}
\end{table}

Our study uses Q\&A posts on \sof, a popular platform used by developers.
Our first step was to find the total number of questions asked about all the ML libraries 
highlighted by some recent surves~\cite{ml16,wang2019various, nguyen2019machine}.
Out of these, we selected 10 popular ML libraries for the study as shown in \tabref{tbl:dataset}. 
%
We excluded the other five libraries because the numbers of questions about 
them were too few (less than 20).

On \sof, each question is rated by the community. 
The score of a question is computed as $S = |N_U| - |N_D|$ where 
$|N_U|$ is the number of upvotes and $|N_D|$ is the number of downvotes.
The higher score is an indicator of the higher quality of the question, which has been used 
in prior works~\cite{meldrum2017crowdsourced}.
\tabref{tbl:dataset} shows the entire distribution of the questions for each library
based on the score $S$.

We selected questions with the score of 5 or higher (bold column in \tabref{tbl:dataset})
to focus on high-quality questions while keeping the workload of manually labeling each question manageable.

Next, we manually classified each \sof question into categories to study them further.
We first discuss the classification of categories and then our labeling process.

%
%

\subsection{Classification of Questions}
\label{subsec:classification}

\begin{figure}[t]
	\includegraphics[width=\columnwidth]{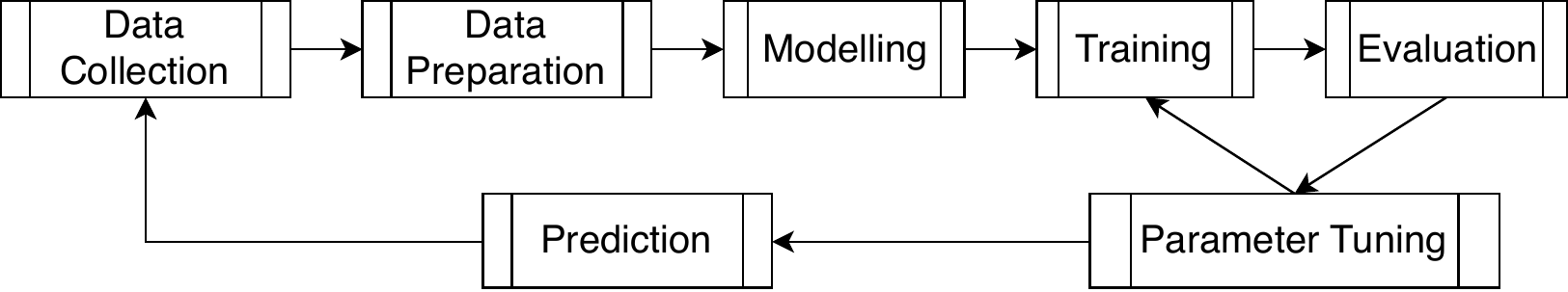}
	\caption{Stages in a typical ML pipeline, based on \cite{stages}.}
	\label{fig:steps}
\end{figure}

\begin{figure*}[t]
	\centering
	\includegraphics[width=0.8\linewidth]{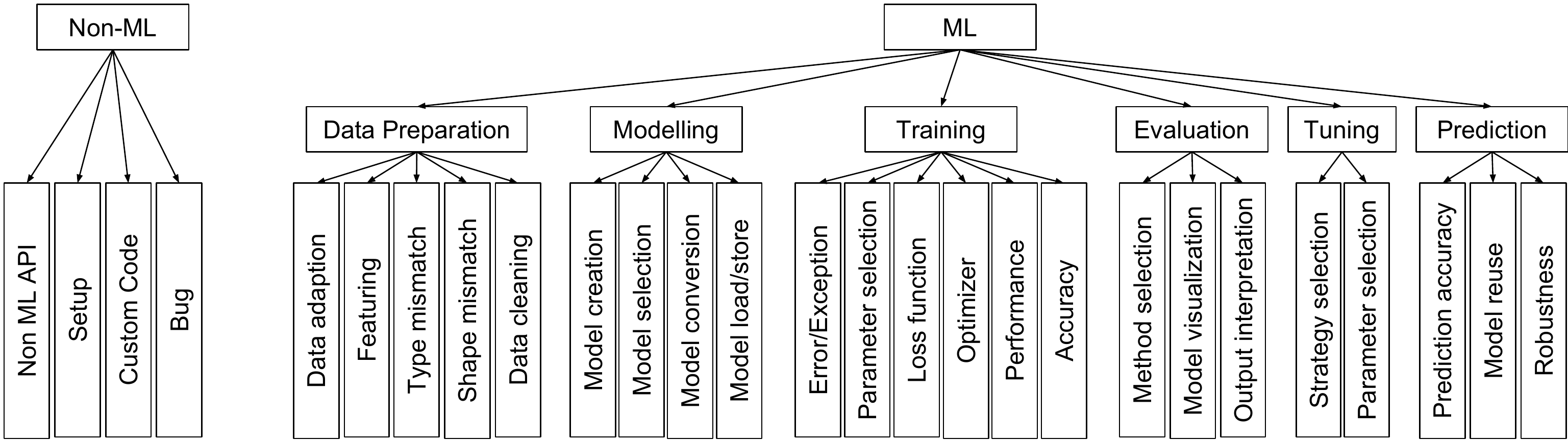}
	\caption{Classification used for categorizing ML library-related Stack Overflow questions for further analysis}
	\label{fig:classes}
\end{figure*}

We classify the questions in \sof into several categories. First, we
classify the questions into two top-level categories based on whether the question is
related to ML or not. Questions related to installation
problems, dependency, platform incompatibility, Non-ML APIs,
overriding the built-in functionality, adding custom functionality fall
into Non-ML category as shown in \figref{fig:classes}.
We classify the ML-related questions into six categories based on the 
stages of a typical ML pipeline~\cite{stages}, also reproduced in \figref{fig:steps}. 
Among those seven stages, data collection is out of the scope of this study because ML libraries do not provide this functionality which leaves us with six categories. 
These six categories are further divided into different sub categories. To find these sub categories one of the Ph.D. (an author of the paper) students with ML expertise first studied 50\% of posts and created the subcategories using open coding scheme adapted from earlier works~\cite{lockyer2004coding,life1994qualitative, strauss1990basics}.
Then these subcategories were sent to three ML experts for review. Based on the review of the experts the subcategories were improved and the process continued until agreement with ML experts was reached. 
The full classification is shown in Figure \ref{fig:classes}.
Next, we describe its categories.


\subsubsection{Data Preparation} 
\label{subsubsec:data-prep}

This top-level category includes questions about converting the raw 
data into the input data format needed by the ML library.


\textbf{Data adaption.\ } Questions under this subcategory are about reading 
raw data into the suitable data format 
required by the library. Data reader provided by the library usually provides this functionality. Questions about converting data, encoding, etc., also fall under 
this subcategory.

\textbf{Featuring.\ } Questions under this subcategory are about feature extraction 
and selection. \emph{Feature extraction} is a process to reduce dimensionality 
of the data where existing features are transformed into a lower
dimensional space. 
\emph{Feature selection} is another strategy of
dimensionality reduction where informative features that have impact on the model are selected. 

\textbf{Type mismatch.\ } Type mismatch happens when the type of data provided 
by the user doesn't match the type required by the ML API. 
For example, if an API needs \code{floating point} data as input but the client 
provides a \code{String} then the ML API raises an exception.

\textbf{Shape mismatch.\ } Shape mismatch occurs when the dimension of the 
tensor or matrix provided by a layer doesn't match the dimension needed by 
the next layer. 
These kinds of errors are very common in deep learning libraries.

\textbf{Data cleaning.\ } Data cleaning phase, sometimes also called data wrangling, includes removal of null values, handling missing values, 
encoding data, etc. 
Without proper data cleaning the training may throw exceptions, 
and accuracy may be suboptimal. 


\subsubsection{Modelling}
\label{subsubsec:choice-model}

The subcategories of this category include:

\textbf{Model selection.\ } This subcategory includes questions related to the choice of the best model and choice of the API version (e.g.
whether to chose SVM or decision tree).

\textbf{Model creation.\ } This subcategory includes questions related to
creating the ML model using the APIs.

\textbf{Model conversion.\ } This subcategory includes questions related to
conversion of a model trained using one library and then using the trained
model for prediction in an environment using another library. For example,
a model trained in \torch can be used for further training or
prediction using \theano.

\textbf{Model load/store.\ } This subcategory contains questions about 
storing models to disk and loading them to use later.

\subsubsection{Training}
\label{subsubsec:training}

The subcategories of this category include:

\textbf{Error/Exception.\ } Questions about errors faced by users in the training phase fall
into this subcategory. The errors may appear due to various reasons. If the errors
are due to shape mismatch or type mismatch we put them into data preparation
category. Otherwise, all errors are placed into this subcategory.

\textbf{Parameter selection.\ } Some frameworks have optional
parameters, and developers have to choose appropriate values for these parameters and also pass relevant values to the required parameters. 
Questions related to these problems fall into this subcategory.

\textbf{Loss function.\ } Questions related to choosing and creating loss
functions fall into this category, e.g., whether to use cosine distance.

\textbf{Optimizer.\ } Questions related to the choice of optimizer are placed into this
subcategory, e.g., should I pick Adam or AdaGrad?


\textbf{Performance.\ } In this subcategory, questions related to long training time and/or 
high memory consumptions are placed.

\textbf{Accuracy.\ } Questions related to training accuracy and/or convergence
are placed into this subcategory.

\subsubsection{Evaluation}
\label{subsubsec:evaluation}

The subcategories of this category include:

\textbf{Evaluation method selection.\ }%
Question related to the problems in the usage of APIs for doing validation 
fall into this subcategory, e.g. ``{\em which of the eight 
APIs for eight different types of validations in \scikit, namely KFold, 
LeaveOneOut, StratifiedKFold, RepeatedStratifiedKFold, RepeatedKFold, 
LeaveOneGroupOut, GroupKFold and ShuffleSplit, should be used?}''



\textbf{Visualizing model learning.\ } The developers sometime need to 
visualize the behavior of the model to get a better understanding of the 
training process and also to know the effects of evaluation on the change of 
loss function and accuracy. Those questions are placed in this subcategory.

\subsubsection{Hyper-parameter Tuning}
\label{subsubsec:hyper-parameter-tuning}

Hyperparameter tuning is used to improve the model's performance. The values of
hyperparameters affect model accuracy. For example, a bad learning rate
may cause a model to learn poorly and give low accuracy.
The subcategories of this category include:

\textbf{Tuning strategy selection.\ }%
Questions about choosing among APIs for different tuning methodologies are 
placed into this subcategory. For example, one poster wondered whether they should use the grid search or randomized search or parameter sampling for parameter tuning in \scikit?


\textbf{Tuning parameter selection.\ }%
This subcategory covers discussions related to the selection of parameters 
for tuning. Some parameters may not have an effect on the model accuracy other 
than increasing the training time while some might have a significant effect on 
the accuracy. For example, the following code from a post is 
trying to tune the kernel and C parameter of the ML
algorithm to find the best combination from values given at line 4. 

\begin{lstlisting}[language=Python,columns=fullflexible]
from sklearn import svm, datasets
from sklearn.model_selection import GridSearchCV
iris = datasets.load_iris()
parameters = {'kernel':('linear','rbf'), 'C':[1,10]}
svc = svm.SVC()
clf = GridSearchCV(svc, parameters)
\end{lstlisting}


\subsubsection{Prediction} 
\label{subsubsec:prediction}

After the model is trained and evaluated, the model is used to predict new input data. 
Questions in this top-level category are about problems faced by the developers during prediction
and include the following subcategories.

\textbf{Prediction accuracy.\ }%
Questions related to prediction accuracy, e.g. due to overfitting,
are placed into this category.

\textbf{Model reuse.\ }%
Developers might have difficulty in reusing existing models 
with their own datasets for prediction to make use of the state of the art 
models from well-known providers. 

\textbf{Robustness.\ }%
Questions in this subcategory are about the stability of the models with 
slight changes, possibly noise, in the datasets. 

\subsection{Manual Labeling}
\label{subsec:manual-labeling}

Manual labeling of the Q\&A dataset was the most important (and time-consuming) step 
before our analysis.
To decrease the bias in manual labeling we recruited three participants. 
Each participant had coursework in both AI and ML and had 
experience using ML libraries to solve problems.
Each participant labeled all the questions producing 9,840 labels.

{\bf Participant Training.\ }
Before the labeling, the participants were provided with the classification shown in 
\figref{fig:classes}. Then, a training session was conducted where each (sub)category 
was discussed and demonstrated using examples. 

{\bf Labelling Efforts.\ } First, each participant gave each question one of the labels from top-level categories namely 
Non-ML, Data Preparation, Modelling, Training, Evaluation,
Tuning, Prediction. Then, (s)he assigned a subcategory. 

We found that, at the steady state, a participant could label around 50-60 
questions per hour. For labeling the whole dataset consisting of 3,283 
questions, each participant took around 1 week time. In total, 168 
person-hours were spent on labeling this dataset.

{\bf Reconciling Results.\ }%
After collecting labels separately from each participant, a moderator then 
compared them. If there was an inconsistency between participants for a 
question, the moderator created an issue in a repository for resolution. 
Among all 3,243 questions, 177 (5\%) needed further discussion. 

Then, the three participants had two in-person 
meetings to discuss those 177 questions. 
The participants read the questions carefully again and voted individually. 
If the votes matched we accepted those as resolved, otherwise 
participants discussed the reasons behind choosing a label
and tried to achieve consensus. 
In most cases, the opinions differed due to the ambiguous nature of the questions. 
For example, for a question asking about suboptimal accuracy, 
it was difficult to say from the question without further exploration whether it is talking about accuracy in the prediction stage or accuracy in the training or evaluation stage. We resolved these type of questions by a careful reanalysis of the Q\&A text.

We measured the inter-rater agreements using Cohen's kappa coefficient ($\kappa$) as shown in \figref{tbl:kappa}. 
It measures the observed level of agreement between raters of a particular 
set of nominal values and corrects for agreements that would appear by chance. 
The interpretation of $\kappa$'s values is shown in \figref{tbl:kappa-interpret}. 
From \figref{tbl:kappa}, we see that the kappa coefficient between all the raters 
involved in the labeling process is more than 0.9 indicating perfect agreements. 
We also computed the \underline{Fleiss coefficient} \cite{hallgren2012computing} which is widely used for finding IRR between more than 2 raters. The Fleiss coefficient was 90.68\% indicating a perfect level of agreement. 
We computed all the IRR coefficients based on the ratings before the discussion for agreement. After the discussion for reconciling the conflicts in the presence of a moderator, the agreement level was 100\%

\begin{figure}%
\centering
\footnotesize
\subfloat[Kappa coefficients ($\kappa$).\label{tbl:kappa}] {
	\begin{tabular}{|l|l | l| l|}
		\hline
		& R1 & R2 & R3 \\
		\hline
		R1 & 1.00 & 0.94 & 0.92 \\
		\hline
		R2 & 0.94 & 1.00 &  0.91 \\
		\hline
		R3 & 0.92 & 0.91 & 1.00 \\
		\hline
	\end{tabular}
}
\subfloat[Interpretation of $\kappa$ value.\label{tbl:kappa-interpret}] {
	\begin{tabular}{|l|l|}
		\hline
		0.00--0.20 & slight agreement \\
		\hline
		0.21--0.40 & fair agreement \\
		\hline
		0.41--0.60 & moderate agreement \\
		\hline
		0.61--0.80 & substantial agreement \\
		\hline
		0.81--1.00 & perfect agreement \\
		\hline
	\end{tabular}
}
\label{fig:reduction}
\caption{Cohen's kappa coefficients for labeling process.}
\end{figure}

\subsection{Threats to Validity}


\textbf{Internal Validity}
In the manual labeling threat can be due to the possibility that labeling could  be biased. We mitigate this threat by using 3 raters and resolving the conflicts via in person meetings. The inter rater reliability coefficient shows that there was perfect level of agreement between raters. 

The possibility of missing relevant posts can also be a threat. We mitigate this bias by collecting the tags that are relevant to a particular ML library. We then collect all the posts containing those tags using \sof API. 

Classification of questions in the top level categories can also pose threat. To mitigate this threat we use the categorization used and described by practitioners and researchers \cite{stages,guzzetta2010machine,meng2016mllib}.  

Classifying the top level categories into subcategories can have bias and missing subcategories due to open coding scheme. To mitigate this threat one PhD student initially studied a subset of posts and came up with the subcategories. Then, three ML experts were consulted and their opinion on the classification was used for multiple round of revisions and improvements.

The ML expertise of the raters can affect the manual labeling. To mitigate this threat we selected raters who have expertise in ML as well as using the libraries in the study. The raters also study the answers and comments in posts to improve their insights.

\textbf{External Validity}
In \sof threat to validity can be low quality posts \cite{rosen2016mobile}, and chronological order of posts. To eliminate the quality threat we studied only the posts that have the tag of the relevant library and then only kept the posts that have score >= 5. This balanced both the quality and labeling efforts.

Chronological order of the posts can introduce threat as some older posts may be resolved in later version of the libraries. To alleviate this threat we classify questions that may appear only due to the API versions into Non-ML category.
 
An external threat can be expertise of the programmers asking the questions. If the questions are asked only by newbies then our results aren't as general.
To understand this threat, we measured {\em reputation}, a metric used by \sof to estimate expertise.
\sof users above 50 are considered reputable and are allowed to comment on posts.
Table \ref{tbl:repu} shows that the mean reputation of programmers asking the questions that we have studied is high indicating that the expertise of the programmers asking the questions is not a threat to our study.

\begin{table}[t]
	\centering
\caption{Reputation of users for posts in our study.}
\setlength{\tabcolsep}{4.8pt}
\begin{tabular}{ |l r r r r r| } 
 \hline
 \rowcolor{gray}
 \textcolor{white}{\bf Library}  & \textcolor{white}{\bf Min}  & \textcolor{white}{\bf Max} & \textcolor{white}{\bf SD} & \textcolor{white}{\bf Mean} & \textcolor{white}{\bf Median}  \\ 
 \hline
 \hline
 
 \caffe  \cite{jia2014caffe} & 28 & 68238 & 7727 & 2252 & 485\\ 
  \rowcolor{lightgray}
 \ho \cite{candel2016deep} & 24 & 7115 &  1635 & 994 & 194 \\ 
\keras \cite{chollet2015keras} & 22 & 68662 & 4639 & 1375 & 354\\ 

  \rowcolor{lightgray}
  \mahout \cite{owen2012mahout} & 26 & 48267 & 7105 & 3217 & 998\\ 
\mllib \cite{meng2016mllib} & 26 & 159552 & 15115 & 3367 & 427\\ 
 
  \rowcolor{lightgray}
  \scikit \cite{pedregosa2011scikit} & 1 & 108807 & 5747 & 1850 & 474\\ 
 
\tensor\cite{abadi2016tensorflow} &  26 & 149231 & 5234 & 1399 & 336\\ 
 
  \rowcolor{lightgray}
  \theano \cite{bergstra2011theano} & 26 & 68692 & 7739 & 2721 & 497\\ 
 
 \torch \cite{collobert2002torch} & 26 & 31205 & 3813 & 1646 & 504\\ 
  \rowcolor{lightgray}
  \weka \cite{holmes1994weka} & 26 & 135428 & 15774 & 4440 & 556\\ 
 Overall  & 1 & 159552 & 6289 & 1764 & 389\\ 
 \hline
\end{tabular}
 \label{tbl:repu}
\end{table}

	\section{Analysis and Results}

\begin{table*}[ht]
	\caption{Percentage of questions in each top-level category across libraries (in \%).}
	\setlength{\tabcolsep}{2.7pt}
	\vspace{-1em}
	\begin{tabular}{ |l r r r r r r r r r r r r r r r| } 
		\hline
		\rowcolor{gray}
		\textcolor{white}{\bf }  & \textcolor{white}{\bf \caffe}  & \textcolor{white}{\bf \ho} & \textcolor{white}{\bf \keras} & \textcolor{white}{\bf \mahout} & \textcolor{white}{\bf \mllib} & \textcolor{white}{\bf \scikit} & \textcolor{white}{\bf \tensor} & \textcolor{white}{\bf \theano}  & \textcolor{white}{\bf \torch} & \textcolor{white}{\bf \weka} & \textcolor{white}{\bf Q1} & \textcolor{white}{\bf Q3}  & \textcolor{white}{\bf IQR} & \textcolor{white}{\bf Median} & \textcolor{white}{\bf SD} \\ 
		\hline
		\hline
		Data preparation  & 14.0 & 41.0 & 16.0 & 17.0 & 33.0 & 26.0 & 16.0 & 17.0 & 23.0 & 30.0 & 16.5 & 29.0 & 12.5 & 20.0 & 8.7\\ 
		
		\rowcolor{lightgray}
		Modelling & 32.0 & 24.0 & 28.0 & $^{\ast}44.0$ & 29.0 & 25.0 & 27.0 & 27.0 & 33.0 & 20.0 & 26.5 & 31.2 & 4.7 & 27.0 & 5.5\\ 
		Training  & 24.0 & 18.0 & 25.0 & 8.0 & 15.0 & 18.0 & 21.0 & 16.0 & 20.0 & 12.0 & 15.0 & 20.7 & 5.7 & 18.0 & 4.7 \\ 
		
		\rowcolor{lightgray}
		Evaluation   & 1.0 & 6.0 & 8.0 & 4.0 & 7.0 & 9.0 & 9.0 & 3.0 & 3.0 & 10.0 & 3.5 & 8.3 & 4.8 & 6.0 & 2.9\\ 
		
		Tuning & 1.0 & $^{\ast}6.0$ & 0.0 & 0.0 & 2.0 & $^{\ast}4.0$ & 1.0 & 1.0 & 0.0 & 0.0 & 0.0 & 1.5 & 1.5 & 1.0 & 1.9\\ 
		
		\rowcolor{lightgray}
		Prediction & 6.0 & 0.0 & 10.0 & 4.0 & 6.0 & 7.0 & 4.0 & 2.0 & 2.0 & 11.0 & 2.6 & 6.6 & 4.0 & 5.0 & 3.2\\ 
		
		Non-ML & 22.0 & 6.0 & 13.0 & 23.0 & 6.0 & 11.0 & 20.0 & 35.0 & 20.0 & 10.0 & 10.3 & 21.4 & 11.1 & 16.0 & 8.6\\ 
		\hline
	\end{tabular}
	
	${\ast}$ indicates the library is an outlier for the category in the corresponding row.
	$IQR = Q3 - Q1$: inter-quartile range.
	SD: standard deviation.
	\label{tbl:cat}
\end{table*}

\begin{table*}[ht]
	\caption{Percentage of questions in each subcategory across libraries (in \%).}
	\setlength{\tabcolsep}{1.3pt}
	\vspace{-1em}
	\begin{tabular}{ |l r r r r r r r r r r r r r r r| } 
		\hline
		\rowcolor{gray}
		\textcolor{white}{\bf }  & \textcolor{white}{\bf \caffe}  & \textcolor{white}{\bf \ho} & \textcolor{white}{\bf \keras} & \textcolor{white}{\bf \mahout} & \textcolor{white}{\bf \mllib} & \textcolor{white}{\bf \scikit} & \textcolor{white}{\bf \tensor} & \textcolor{white}{\bf \theano}  & \textcolor{white}{\bf \torch} & \textcolor{white}{\bf \weka} & \textcolor{white}{\bf Q1} & \textcolor{white}{\bf Q3} & \textcolor{white}{\bf IQR}  &  \textcolor{white}{\bf Median} & \textcolor{white}{\bf SD} \\ 
		\hline
		\hline
		Data adaptation & 9.84 & 35.29 & 7.90 & 14.58 & 25.2 & 8.64 & 9.22 & 10.41 & 22.95 & 20.51 & 9.37 & 22.3 & 12.93 & 12.5 & 8.70 \\ 
		\rowcolor{lightgray}
		Featuring & 0 & 5.88 & 1.09 & 0 & 4.20 & 9.34 & 0.74 & 0.52 & 0 & 4.27 & 0.13 & 4.3 & 4.17 & 0.92 & 3.03 \\ 
		Type mismatch & 1.52 & 0 & 1.09 & 0 & 2.52 & 2.92 & 2.02 & 2.08 & 0 & 1.71 & 0.27 & 2.07 & 1.80 & 1.61 & 1.02 \\ 
		\rowcolor{lightgray}
		Shape mismatch & 1.52 & 0 & \textbf{$^{\ast}5.50$} & 0 & 0 & 1.86 & 2.62 & 2.08 & 0 & 0 & 0 & 2.03 & 2.03 & 0.75 & 1.70 \\ 
		Data Cleaning  & 1.52 & 0 & 0.55 & 2.10 & 2.52 & 3.62 & 2.09 & 1.60 & 0 & 3.41 & 0.79 & 2.40 & 1.61 & 1.82 & 1.22 \\ 
		\hline
		\hline
		\rowcolor{lightgray}
		Model creation & 26.52 & 17.64 & 25.88 & \textbf{$^{\ast}43.75$} & 23.52 & 21.37 & 23.01 & 23.43 & 22.95 & 21.36 & 21.77 & 25.30 & 3.53 & 23.22 & 6.70 \\ 
		Model selection  & 0 & 0 & 0.55 & 0 & 0.84 & \textbf{$^{\ast}2.10$} & 0.60 & 0 & 0 & 0 & 0 & 0.58 & 0.58 & 0 & 0.64 \\ 
		\rowcolor{lightgray}
		Model conversion  & 3.79 & 0 & 0.27 & 0 & 0.84 & 0.33 & 2.25 & 2.60 &  4.91 & 3.41 & 0.24 & 3.21 & 2.97 & 1.54 & 1.71 \\ 
		Model load/store & 1.50 & 5.88 & 1.63 & 0 & 5.04 & 1.75 & 1.94 & 1.01 & 4.91 & 1.71 & 1.55 & 4.20 & 2.65 & 1.73 & 1.88 \\ 
		\rowcolor{lightgray}
		\hline
		\hline
		Error/Exception  & 0.76 & 5.88 & 5.50 & 4.20 & 5.88 & 4.78 & 5.32 & 5.72 & 1.64 & 2.56 & 2.96 & 5.67 & 2.71 & 5.10 & 1.80 \\ 
		Parameter selction & 9.10 & 5.88 & 5.50 & 0 & 2.52 & 3.97 & 3.74 & 2.60 & 8.20 & 5.13 & 2.89 & 5.78 & 2.89 & 4.50 & 2.60 \\ 
		\rowcolor{lightgray}
		Loss function & 6.10 & 0 & 4.09 & 0 & 0.84 & 1.40 & 3.74 & 2.60 & 3.30 & 1.71 & 0.98 & 3.60 & 2.62 & 2.16 & 1.86 \\ 
		Optimizer & 2.30 & 0 & 1.09 & 2.10 & 0 & 0.70 & 2.77 & 1.04 & 3.30 & 0.85 & 0.74 & 2.20 & 1.46 & 1.07 & 1.07 \\ 
		\rowcolor{lightgray}
		Performance & 2.30 & 5.88 & 6.27 & 2.10 & 5.04 & 3.27 & 4.87 & 3.12 & 1.64 & 0.85 & 2.13 & 5.00 & 2.87 & 3.20 & 1.80 \\ 
		Accuracy & 3.78 & 0 & 2.45 & 0 & 0.84 & 3.62 & 0.90 & 1.04 & 1.64 & 0.85 & 0.84 & 2.30 & 1.46 & 0.97 & 1.30 \\ 
		\hline
		\hline
		\rowcolor{lightgray}
		Eval. strategy selection & 0.75 & 0 & 2.18 & 2.08 & 5.04 & 3.85 & 5.24 & 0 & 1.64 & 8.54 & 0.84 & 4.7 & 3.86 & 1.86 & 2.64 \\ 
		Visualization & 0 & 0 & 1.63 & 0 & 0 & 2.68 & 1.65 & 0 & 0 & 2.68 & 0 & 1.23 & 1.23 & 0 & 0.95 \\ 
		\rowcolor{lightgray}
		Output interpretation & 0 & \textbf{$^{\ast}5.88$} & \textbf{$^{\ast}3.82$} & 2.1 & 1.68 & 2.21 & 2.17 & 2.60 & 1.64 & 1.71 & 1.69 & 2.50 & 0.81 & 2.13 & 1.47 \\ 
		\hline
		\hline
		Tuning strategy selection & 0.75 & \textbf{$^{\ast}5.88$} & 0.27 & 0 & 1.68 & 3.50 & 0.45 & 1.04 & 0 & 0 & 0.07 & 1.52 & 1.45 & 0.60 & 1.82 \\ 
		\rowcolor{lightgray}
		Tuning param. selection &0 & 0 & 0 & 0 & 0 & \textbf{$^{\ast}0.81$} & 0.08 & 0 & 0 & 0 & 0 & 0 & 0 & 0 & 0.24 \\ 
		\hline
		\hline 
		Prediction accuracy & 6.10 & 0 & 6.81 & 4.20 & 5.04 & 5.25 & 3.97 & 2.08 & 1.64 & 8.54 & 2.55 & 5.85 & 3.30 & 4.60 & 2.44 \\ 
		\rowcolor{lightgray}
		Model reuse & 0 & 0 & \textbf{$^{\ast}1.37$} & 0 & 0 & 0.23 & 0.22 & 0 & 0 & \textbf{$^{\ast}1.71$} & 0 & 0.23 & 0.23 & 0 & 0.60 \\ 
		Robustness & 0 & 0 & 1.65 & 0 & 0.84 & 1.28 & 0.30 & 0 & 0 & 0.85 & 0 & 0.85 & 0.85 & 0.15 & 0.59 \\ 
		\hline
		\hline
		\rowcolor{lightgray}
		Non-ML API & 2.27 & 0 & 2.72 & 4.20 & 2.52 & 2.80 & 4.40 & 2.08 & 3.27 & 1.71 & 2.13 & 3.15 & 1.02 & 2.63 & 1.19 \\ 
		Setup & 16.67 & 5.88 & 9.53 & 18.75 & 2.52 & 5.95 & 14.50 & 30.72 & 16.39 & 7.69 & 6.39 & 16.60 & 10.21 & 12.05 & 7.90 \\ 
		\rowcolor{lightgray}
		Custom code & 1.52 & 0 & 0.81 & 0 & 0.84 & 1.51 & 1.05 & 1.56 & 0 & 0.85 & 0.21 & 1.40 & 1.19 & 0.84 & 0.60 \\ 
		Bug & \textbf{$^{\ast}1.52$} & 0 & 0 & 0 & 0 & \textbf{$^{\ast}0.23$} & 0.07 & 0 & 0 & 0 & 0 & 0.06 & 0.06 & 0 & 0.45 \\ 
		\hline
	\end{tabular}
	
	${\ast}$ indicates the library is an outlier for the subcategory in the corresponding row.
	$IQR$ and $SD$ are defined in \tabref{tbl:cat}.
	\label{tbl:subcat}
\end{table*}
We have proposed four research questions to understand what developers ask about ML. We explore the answers to these research questions in this study. The research questions cover the following aspects:
identifying the difficult stages in the current ML pipeline faced by the developers (\textbf{RQ1}), understanding whether the problems faced by the developers are only due to the design of library or there are some problems inherent to ML (\textbf{RQ2}), exploring whether some of the libraries are more difficult in certain stages and are there libraries that shows comparable difficulties in all the stages (\textbf{RQ3}), exploring whether the problems faced by the developers changed over time or they stayed consistent (\textbf{RQ4}).
Next, we answer these questions using a statistical analysis summarized in Tables \ref{tbl:cat} and  \ref{tbl:subcat} and present our findings.
Our raters have read and agree with these findings.

\section{RQ1: Difficult stages}
If we know the relative difficulty of ML stages for developers, 
then software engineering R\&D and educational efforts can prioritize
work on challenging stages. This section explores this question. 
\subsection{Most difficult stage}
As \tabref{tbl:cat} shows, and as expected the model creation is the most difficult, but surprisingly data preparation is the next difficult stage which turns out to be more difficult than training stage.

Model creation has the median of 23\% across all the libraries which is the highest compared to all other stages in the ML pipeline. Some of the libraries for distributed ML like \mahout, \torch, \caffe, \mllib have abnormally high difficulty in model creation stage. This suggests that machine learning in distributed environment is not developer friendly yet. 
\findings{Model creation is the most challenging (yet critical) in ML 
	pipeline, especially for libraries supporting distributed ML on 
	clusters like \mahout and \mllib. 
}{This calls for tool support in creating models, 
	especially in distributed machine learning.}  
Finding 1 provides indication that, tool support in creating models, 
especially in distributed machine learning is needed.
Enhancing tool support to make model creation in distributed environment 
easier and research on detecting and resolving problems in model creation 
is needed. 

Further analysis showed that libraries that model creation is 
especially harder for ML libraries that require developers to use multiple 
configuration languages to configure their models, for example in \caffe.
For example in \caffe, questions about using multiple languages are discussed frequently.
According to a case study by Amershi \etal \cite{amershi2019software}, Microsoft developers face similar issues in the machine learning pipeline. It says that though Data Availability, Collection, Cleaning, and Management are most challenging for all three groups of the developer but Model Evolution, Evaluation, and Deployment are more significant for all groups according to the frequency. Our results from studying the posts support what was known at a corporate community and have found several new findings.



\subsection{Data preparation}
This top level category includes questions about adapting the data to the format required 
by the library, featuring, dealing with type and shape mismatches, and data cleaning.
All together, this stage is the next most difficult stage across ML libraries
(median 20\%). 

\findings{Data preparation, especially data adaptation, is the second most difficult stage in ML pipeline.}{Data preparation tools are needed and libraries need strong support for data cleaning and verification.}

Further analysis showed that ML libraries that use uncommon formats lead to additional 
difficulties among the developers to understand the format, use the new formats in their
software, data wrangling and preprocessing to the format of the data. 
For example, \weka, \mllib have higher problem in data adaption due to their use of 
uncommon ARFF and RDD formats of data.
For some libraries data preparation turns out to be ever more challenging compared to 
model creation. For example, \ho, \torch and \weka have 35.29\%, 22.95\% and 20.51\% 
of posts, respectively, about data adaptation.

This finding suggests that the tradeoff in the design of data preparation APIs, 
e.g. use of custom formats, needs more study. Interestingly, most of the ML textbooks 
and courses spend little time on data preparation related discussions. 




Surprisingly Tuning and Prediction stages of the ML pipeline---topics discussed frequently 
in the ML research papers---appear infrequently in \sof questions.

\section{RQ2: Nature of problems}

Are some difficulties inherent to ML and thus 
all ML libraries face them? If so, general solutions could be developed
and adapted to all ML libraries. Otherwise, design of the specific library
could be improved by utilizing lessons learned in this section.  

\subsection{Type mismatch}

Type mismatch questions have median of 1.61\%, SD of 1.02\%, and IQR of 1.80\%. 
The smaller IQR indicates that type mismatch appears in most of the ML
libraries. \scikit, \mllib, \theano and \tensor  have higher difficulties in 
type-related problems with 2.92\%, 2.52\%, 2.08\% and 2.02\%, respectively. 
\mllib uses a custom data format called RDD that seems to make type-related 
problems more frequent for this library. There are also questions about failures 
due to type mismatch in \scikit, \tensor and \theano as their APIs have type 
requirements that are not currently checked. 

\findings{Type mismatches appear in most ML libraries.}{Type checkers are 
	desirable for ML libraries.}
The finding suggests that ML libraries have not focused on type correctness
and ML-specific type correctness. 
A static analysis tool might be 
able to prevent the majority of these problems.
To understand the characteristics of the type mismatch related posts, we randomly select 44 \sof posts. We found 31 out of 44 problems were caused by the abstraction created by the libraries to create ML types. The other 13 were standard Python type errors. 
As an example, the following exception is thrown due to an ML type error.
\begin{lstlisting}[language=Python,columns=fullflexible]
ValueError: ('Unknown loss function', ':root_mean_squared_error')
\end{lstlisting}
\subsection{Shape mismatch}

\begin{figure}[h!t]
	\includegraphics[width=0.8\columnwidth]{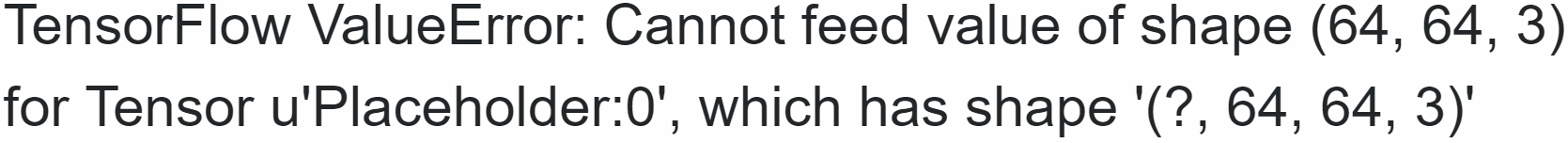}
	\caption{\href{https://stackoverflow.com/questions/40430186}{Question 40430186}: An example showing dimension or shape mismatch problem in training in ML.}
	\label{fig:shape}
\end{figure}

Shape mismatch related questions have median of 0.75\%, SD of 1.70\%, and IQR 
of 2.03\%. This problem appears in all deep learning library in which \keras 
is an outlier with 5.50\%. In these libraries, shapes of neurons at adjacent 
layers must be compatible otherwise the library will throw exceptions during 
training or fail during prediction. 
An example is shown in \figref{fig:shape}.

\findings{Shape mismatch problems appear frequently in deep 
	learning libraries. \keras is an outlier in this subcategory with 5.5\% of 
	posts.}{Tool support for verifying shape and dimension compatibility
	is needed for deep learning libraries. Dependency of data on 
	model architecture needs to be verified and dynamic modification of the 
	network as per data shape may be needed.}

The finding suggests that techniques for verifying shape and dimension compatibility
are needed for deep learning libraries. 
Such techniques could verify if the data conforms to 
model architecture, and dynamic modification of the 
network against data shape.

Abstract APIs that hide the details of inner-working of the deep learning networks can 
further complicate matters. 
To illustrate consider the following \keras code.

\begin{lstlisting}[language=Python,columns=fullflexible]
def CreateModel(shape):
	if not shape:
		raise ValueError('Invalid shape')	
	logging.info('Creating model')
	model = Sequential()
	model.add(LSTM(4, input_shape=(31, 3)))
	model.add(Dense(1))
	model.compile(loss=`mean_squared_error', optimizer=`adam')
	return model

\end{lstlisting}

The error is at line 6 where an invalid value of $(31,3)$ is passed 
to $input\_shape$. The accepted answer suggests that $input\_shape$ should be 
$(32,1)$ instead. The user could not verify statically whether the built 
model has compatible shape or if there are any unconnected or extra ports while 
building the model. If we had the tools that could tell the developer 
that using dimension (32,1) can cause 2 out of 3 ports of the next layer to 
be unconnected then it would be much easier for the developer to find these 
errors by themselves. These kind of errors could be detected by program analyses 
and by providing feedback to the users. In fact, many discussions in these 
high-scored posts call for richer analysis features.
To understand the reason behind the \keras being an outliner in Shape Mismatch sbcategory, we have selected 60 random posts from the dataset. We have found that 21 out of 60 shape mismatch problems are from Keras and the shape mismatch in \keras occurs due to the abstraction of APIs used to create layers in the network. The dimension of the layers violate the contracts between the layers without giving any hints to the developer. 
\subsection{Data Cleaning}

As shown in \tabref{tbl:subcat}, data cleaning related questions across the 
libraries have median of 1.82\%, SD of 1.22\% and IQR of 1.61\%. Most of the 
libraries have questions about data cleaning stage except for \ho and \torch.
This is not surprising since  
data cleaning is an integral part of any data science pipelines. 
Libraries \scikit, \weka and \mllib have the most questions. 

\findings{Most libraries have problems in data cleaning.}{ Data cleaning 
	needs verification tools to ensure that all required steps in the process are performed. 
}
This finding suggests that tool support for data cleaning
is needed, but such techniques
may need to overcome inherent technical challenges. 
The abstract APIs in these libraries sometimes make 
cleaning fail. For example, the \code{nan} values in the \code{dataframe} 
needs to be converted first into numpy \code{nan} type before they can be 
cleaned using APIs provided by \scikit. 
Furthermore, these failures do not clearly indicate the root cause
making diagnostics difficult.
%

\subsection{Model creation}
\begin{figure}[h!t]
	\includegraphics[width=0.8\columnwidth]{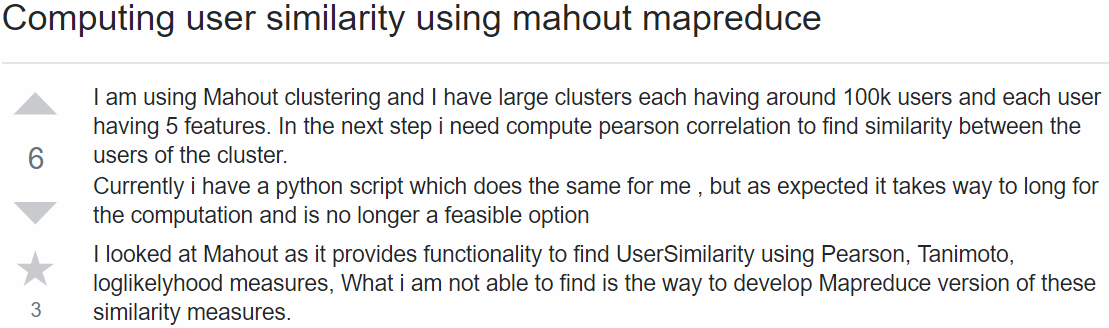}
	\caption{\href{https://stackoverflow.com/questions/12319454}{Question 12319454}: An example question on model creation for distributed ML using \mahout.}
	\label{fig:mahout}
\end{figure}

In model creation subcategory, {\em the most difficult stage according to RQ1}, 
we see problems that are both inherent to ML,
and specific to design choices in the library. 
Inherent difficulty of distributed ML is a major source of questions, e.g. 
see \figref{fig:mahout}.

%
%

Deep learning libraries like \caffe, \keras, \theano and \tensor also have 
higher percentages of questions about model creation with 26.52\%, 25.88\%, 23.43\% and 23.01\%, respectively. This shows that model creation 
for deep neural networks is difficult as well. 



When we study the questions about \caffe we see that \caffe users have 
problems in model creation due to the  dependency of the model on multiple 
files. To create a model successfully, one  needs to make a schema file in 
protobuf format, create a solver file and write code in C++ or Python to 
build the model~\cite{caffetutorail}. 
Having several components complicates matters.
In our study, 36 out of 135 questions about \caffe are about model creation problems. 



\subsection{Error/Exception}

Error/Exception subcategory has the median of 5.10\%, SD of 1.80\% and IQR of 2.71\%. 
All the libraries have issues on runtime error/exception. 
Surprisingly, though model creation seems problematic in \caffe, 
runtime failure is very low in \caffe with 0.76\%. 
\mllib, \ho, \keras, \tensor and \scikit have higher percentage of runtime errors with 5.88\% and 5.88\%, 5.50\%, 5.32\% and 4.78\%, respectively.

\findings{
	Questions on exceptions/errors are prevalent.
}{
	Deep learning and distributed ML libraries \change{have more posts about runtime error at training time}. 
	This indicates static and dynamic analysis tools are needed for such 
	libraries.}

\begin{figure}[h!t]
	\includegraphics[width=0.8\columnwidth]{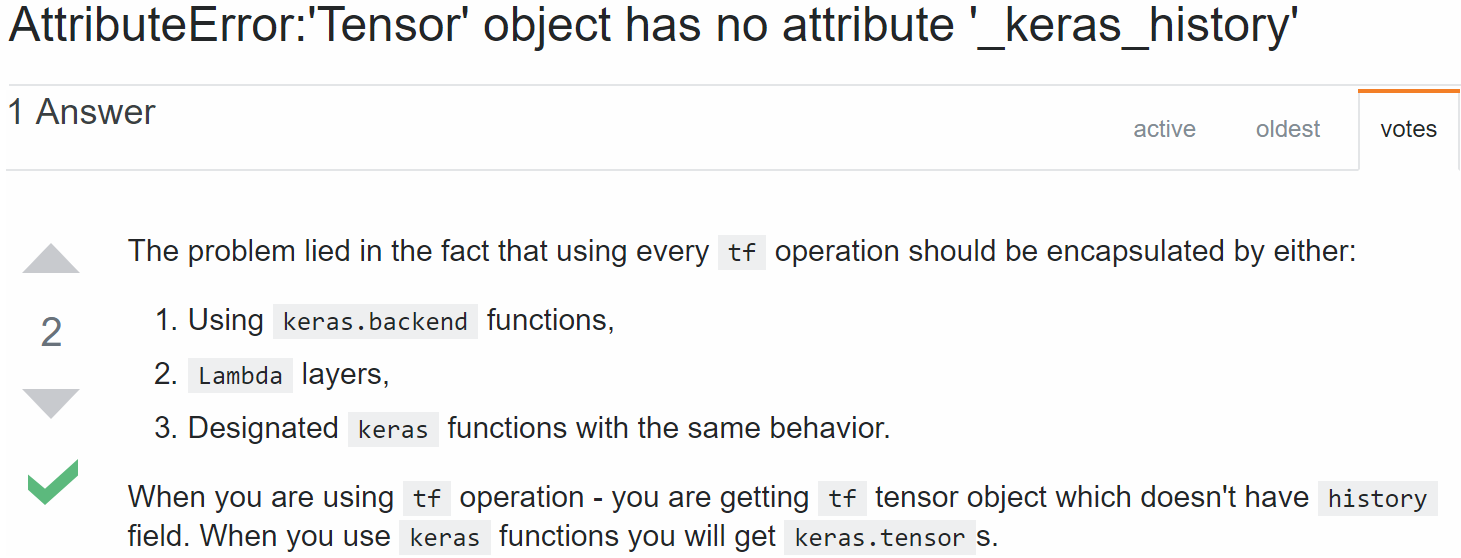}
	\caption{\href{https://stackoverflow.com/questions/45030966}{Question 45030966}: An example question about \keras showing abstraction in deep learning libraries could make identifying root cause of an error/exception difficult.}	\label{fig:ob6}
\end{figure}

This finding suggests that debugging and 
monitoring facilities for ML needs much improvement
to help developers resolve error/exception independently.
We dug deeper to determine where debugging and monitoring might be most helpful
and found that deep learning and distributed ML libraries 
have more posts about runtime errors at training time, e.g. when a model 
is throwing an exception at training time, a model is not converging or 
learning as the iteration of training goes on, a model is not predicting 
well, etc.
Fortunately, some 
recent work has started to address these issues~\cite{chakarov2016debugging,tdebug}, 
but much more work is needed. Due to the lack of debugging 
tools to monitor pipelines causes of failure are hard to identify. 
More abstract deep learning libraries 
throw more runtime exception during training, e.g. see Figure \ref{fig:ob6}.

%

\subsection{Parameter selection}

We expected parameter selection to be an inherent ML issue but found some 
variation between libraries, median of 4.50\%, SD of 2.60\% and 
IQR of 2.89\%, suggesting key differences among libraries. 
\caffe and \torch have comparatively more problems with 9.10\% and 8.20\%, 
respectively. Libraries like \keras , \weka, \ho, \mllib shows larger 
percentage of questions on choice of parameters. 

\findings{
	Parameter selection can be difficult in all the ML libraries.
}{
	Meta-heuristic strategies can be helpful.}
For selecting parameters adding support for meta-heuristic strategies 
in the libraries can be helpful.

\subsection{Loss function selection}

Loss functions are used to quantify the difference between values 
predicted by the model and actual values (labels). 
Our results shows that developers have difficulty selecting an appropriate 
loss function but the extent of difficulties varies across the libraries
(median of 2.16\%, SD of 1.86\% and IQR of 2.62\%).
All deep learning libraries have comparatively more questions about loss function,
for example \caffe, \keras, \tensor and \torch have the 
highest percentages of 6.10\%, 4.09\%, 3.74\% and 3.30\%.

\findings{Choice of loss function is difficult in deep learning libraries.}
{This indicates the necessity to develop automatic suggestion algorithms and tools 
	for selecting function for deep learning.}

This indicates the necessity of further research on the usage of loss function in 
deep learning libraries, e.g. on loss function recommendation. The selection of the loss function is primarily dependent on the type of the problem. A wrong selection of the loss function can cause a machine learning model to perform poorer (low accuracy) or can decrease the security of a model by decreasing the robustness that can be utilized by attackers to perform adversarial attack\cite{saito2018effects}.

\subsection{Training accuracy}

We expected training accuracy to be an inherent ML issue impacting all 
libraries; however, there are few questions about this on \sof.
\caffe and \scikit stood out with 3.78\% and 3.62\% questions about training accuracy.
These libraries provide highly abstract APIs and a large number of optional parameters
that need to be selected. 

\findings{Abstract ML libraries have higher percentage of questions about 
	training time accuracy and convergence.}{This 
	indicates the necessity to embed dynamic analysis of learning behavior along 
	with abstract APIs and the development of these analysis tools.}

This suggests that the library documentation could be clearer about the impact
of optional parameters on training accuracy. Secondly, recommendation system
could be developed for parameter recommendation based on dynamic traces.

\subsection{Tuning parameter selection}

Like accuracy, we considered tuning parameter selection to be an inherent ML issue,
impacting those libraries more that have higher number of parameters. 
Even though not too many libraries have questions about it, \scikit and \tensor
stand out. \tensor has higher usage and questions in general, but \scikit was 
as expected due to the large number of optional parameters.

%
%
%
As an example, consider creating \texttt{AdaBoostClassifier} 
with 5 optional parameters initialized to some default values shown below. 

\begin{lstlisting}[language=Python]
class sklearn.ensemble.AdaBoostClassifier(
base_estimator=None, n_estimators=50, learning_rate=1.0, algorithm='SAMME.R', random_state=None)
\end{lstlisting} 

\begin{figure}[t]
	\includegraphics[width=0.8\columnwidth]{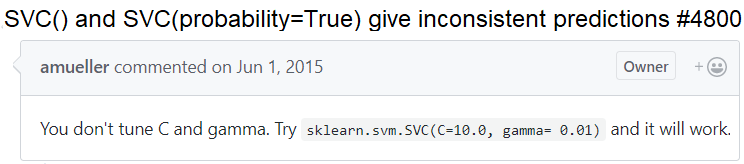}
	\caption{\href{https://github.com/scikit-learn/scikit-learn/issues/4800}{\scikit issue \#4800}: An example of hyperparameter tuning problem. The user filed a bug report, but a developer of the library responded that the problem was with hyperparameter tuning.}
	\label{fig:tune}
\end{figure}

The \texttt{base\_estimator} is set to \texttt{None} but the user may need to choose an 
estimator to get the best performance. Learning rate is by default set to 1.0. %
At this learning rate, it is highly likely that the model will not learn 
anything. So the user may often use these APIs incorrectly and wonder why ML 
model is not producing useful results. Since these are optional 
parameters, the user will not even get any error or warning. Finding good 
values for these parameters and tuning them to make the best model, avoiding 
over-fitting are frequent questions among developers using \scikit. There 
have been some GitHub issues filed to the repository of \scikit as bugs (See 
\figref{fig:tune} for an example) but the underlying problem was that the 
developer was not able to trace why the model is not showing expected 
accuracy, and unable to tune hyperparameters. We have found 36 questions out 
of 849 in \scikit asking help about hyperparameter tuning.

\findings{\scikit has more difficulty in hyper parameter tuning compared to 
	other libraries}{Due to the presence of a lot of optional parameters in the 
	APIs, the tuning of these parameters is more problematic than other libraries. 
	This gives intuition of having tools to suggest parameters that may have 
	effect on a particular model and particular problem. }

Overall, our results from this and two previous subsections suggest that 
{\em parameter recommendation is an urgent need for ML libraries, especially those
that have a lot of optional parameters}.

\subsection{Correlation between libraries} 

\begin{figure}[ht]
	\includegraphics[width=\columnwidth]{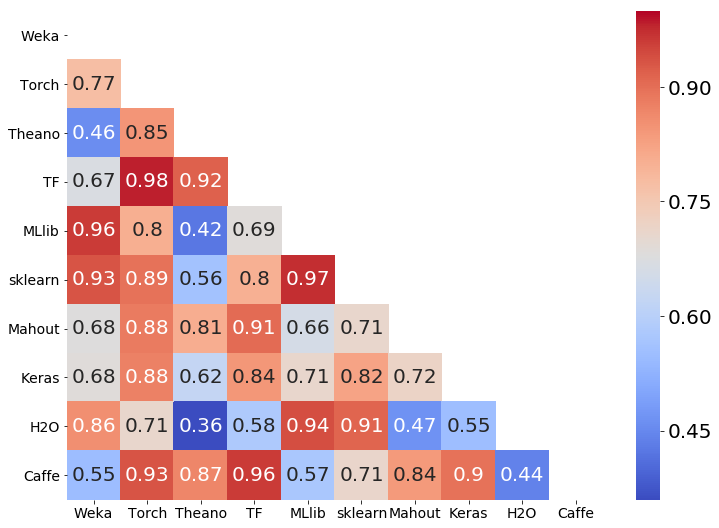}
	\caption{Correlation between distributions of percentage of questions over stages of the libraries.}
	\label{fig:corrlib}
\end{figure}

Next, we study whether the pattern of problems exhibited by libraries have  
similarities. 
The correlation between libraries based on common pattern of problems is shown in \figref{fig:corrlib}. 
We have identified two major groups.

\textbf{Group 1.\ }  \weka,~\ho,~\scikit,~and~\mllib form a strongly correlated group with correlation coefficient greater than 0.84 between the pairs. This suggests that the problems appearing in these libraries have some correlation and the difficulties of one library can be described by the difficulty of other libraries in the group.

\findings{\weka,~\ho,~\scikit,~\mllib form a strong correlated group with correlation coefficient greater than 0.84 between the pairs indicating that these libraries have similar problem in all the ML stages.}
{Analysis framework for one library can be reused in other library in this group.}

This finding is interesting because other than \ho, other libraries in this category don't support deep learning.
We believe that the correlation may be because each of these libraries support many different ML algorithms and allow the user to select an algorithm for their tasks.
This design choice is markedly different from the other group that are specialized for a single ML algorithm.

\textbf{Group 2.\ } \torch,~\keras,~\theano,~and~\tensor form another group with strong correlation of more than 0.86 between the pairs. These libraries are all specialized for deep learning. 

\findings{Deep learning libraries \torch,~\keras,~\theano~and~\tensor form another group with strong correlation of more than 0.86 between the pairs indicating these libraries follow similar problem in all the stages}{Problems of deep learning libraries currently follow a common pattern which can be leveraged by developing analysis framework for one library and studying how much solution is transferable from one library to another library.}

This finding is interesting because each of these deep learning libraries have adopted different design and philosophies. \tensor and \torch are focused on providing low-level general facilities, \keras focuses on high-level abstractions, whereas \theano focuses on efficiency on both CPU and GPU.
Our finding suggests that despite different design philosophies followed by each of these ML libraries, the problems are interrelated for the libraries in this category. So, the software engineering research results for one library may generalize to other deep learning libraries.


\subsection{API Misuses in All ML Stages}

The ML libraries have APIs that are very often misused. 
To identify the misuses we have studied both the questions asked by some developer and the well accepted answers. If the answers pointed out to incorrect or wrong use of API and provided solution using correct use of APIs, we marked them as posts containing API misuse.
API misuse is seen across all the stages of ML pipeline. 

\begin{figure}[t]
	\centering
	\subfloat[Question] {
		\includegraphics[width=\linewidth]{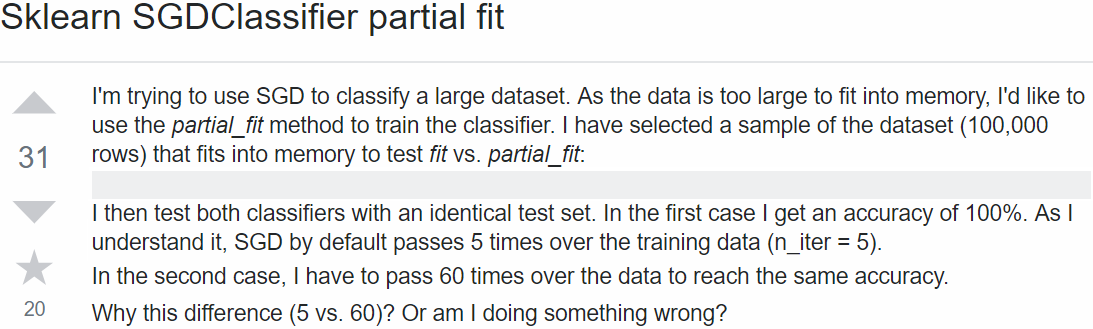}
	}\\	
	\subfloat[Best accepted answer] {
		\includegraphics[width=\linewidth]{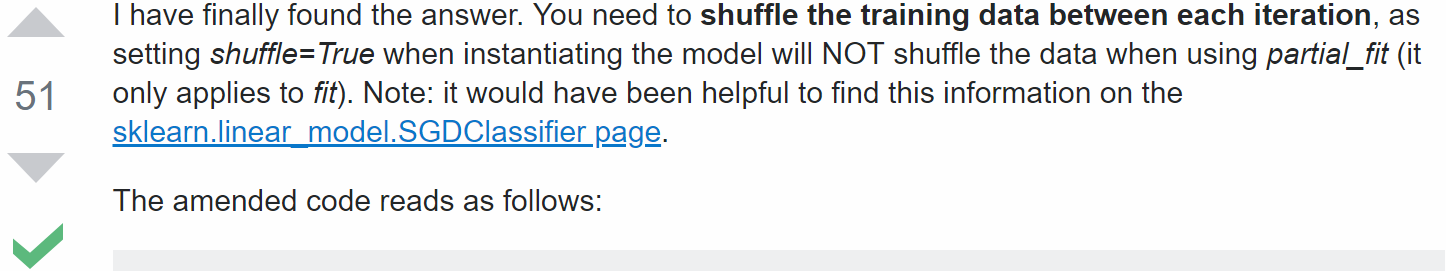}
	}
	\caption{\href{https://stackoverflow.com/questions/24617356}{Question 24617356}: An example showing the API misuse problem in ML libraries. Code snippets are omitted.}
	\label{fig:apimis}
\end{figure}

For example, see Figure \ref{fig:apimis} where a user is asking
that their training takes much time or longer number of iterations to get a 
certain training accuracy. 
When they use one API they are able to achieve the desired accuracy  
in 5 iterations where in the other API they need 60 iterations to reach 
the same accuracy. The second API works fine, without any error and 
eventually reaches the same accuracy. But still, the user is puzzled 
that almost 12 times higher number of iterations are required when 
using the second API. 
The answer in Figure \ref{fig:apimis}b suggests that the second API needs the 
data to be shuffled properly before passing to the API in every iteration. 
Making that change solves the performance problem.
This is an example of 
API misuse where the precondition of the second API is not satisfied which 
leads to a performance bottleneck.
For another example, let's consider a problem related to the creation of a
NaiveBayes model. Only a part of the code snippet where API misuse occurred is
shown below:

\begin{lstlisting}[language=Python,columns=fullflexible]
def convert_to_csr_matrix(vectors):
	logger.info("building the csr_sparse matrix representing tf-idf")
	row = [[i] * len(v) for i, v in enumerate(vectors)]
	row = list(chain(*row))
	column = [j for j, _ in chain(*vectors)]
	data = [d for _, d in chain(*vectors)]
	return csr_matrix((data, (row, column))) 
\end{lstlisting}

The code failed to work successfully giving dimension mismatch error in 
some parts of the code. 
The solution to the problem is to properly use the API $csr\_matrix()$. 
This API needs to have a shape parameter defined explicitly and the correct 
way to use the API is to explicitly define the shape shown in the code below.

\begin{lstlisting}[language=Python,columns=fullflexible]
return csr_matrix((data, (row, column)), shape=(len(vectors), dimension))
\end{lstlisting}

We have observed another kind of API misuse due to API update by the 
library provider. 
To illustrate, consider the code below that worked well in Apache Spark 
\mllib version $<$ 2.0. For Apache Spark version $>=$ 2.0, this API doesn't work.
This is one of the top voted questions on Apache Spark \mllib category. 

\begin{lstlisting}[language=Python,columns=fullflexible]
from pyspark.mllib.clustering import KMeans
spark_df = sqlContext.createDataFrame(pandas_df)
rdd = spark_df.map(lambda data: Vectors.dense([float(c) for c in data]))
mdl = KMeans.train(rdd, 2, maxIterations=10, runs=30, initializationMode="random")
\end{lstlisting}

\mllib version 2.0 isn't backward compatible and so the code at Line 3 
is outdated and must be replaced by the following 

\begin{lstlisting}[language=Python,columns=fullflexible]
rdd = spark_df.rdd.map(lambda data: Vectors.dense([float(c) for c in data]))
\end{lstlisting}

We have found that similar version incompatibility problems are also 
prevalent in other ML libraries.

Besides, the API misuse scenarios discussed above, many other kinds of API 
misuse are common in ML libraries, and a more detailed analysis and 
categorization of errors is needed (much like MUBench~\cite{Amann2016}). Some 
common problems include failure to find important features, improperly 
preparing the dataset, performance, over-fitting problems, suboptimal 
prediction performance, etc. A detailed analysis of API misuse is beyond the 
scope of this work.

\section{RQ3: Nature of libraries}
In this section we explore whether some of the libraries are more 
difficult in certain stages and are there libraries that shows 
comparable difficulties in all the stages (\textbf{RQ3}).
To answer RQ3, we look at three measures. Which libraries have non-zero percentage of
questions under the majority of subcategories? Which libraries have above median 
percentage of questions under the majority of subcategories?
Which libraries have outliers?

It turns out that \scikit and \tensor have questions under all subcategories, 
and \keras, \weka, \mllib, \caffe, and \theano have questions under the majority 
of subcategories. On the other hand, \ho, \mahout and \torch have questions concentrated 
under few subcategories and other subcategories have no questions.
We further observed subcategories under which \ho have the majority of questions
and found that the majority of the questions are in the initial stages such as 
how to adapt data to use within \ho, how to create a model, or how to setup 
to use the library adequately.
We also observed similar trends for \mahout except it has proportionally higher
percentage of questions about model creation and setup.

\findings{Early stages for \ho and \mahout especially setup and model creation
	have comparatively higher percentage of questions compared to later stages.
}{
	Model creation in parallel machine learning libraries is difficult. 
}

This may suggest that getting started is harder with \ho and \mahout.
Reflecting further on the nature of \ho and \mahout, there is a key similarity 
between the two libraries. Both present non-traditional models of computation 
to the developers. \ho presents a workflow like model, and \mahout is for 
distributed ML.
The absence of questions for later stage subcategories might suggest 
either that developers who started with \ho and \mahout stopped using the 
library or that all developers who faced problems getting started with \ho and \mahout
continued using the library without any major difficulties, and had no questions.
Further research is needed to understand which was the case and we didn't find
any definitive evidence during this study to suggest either way.

Next, we look at libraries that have above median percentage of questions under the
majority of subcategories. At the top, $>$50\% subcategories, are 
\tensor (20 subcategories), \scikit (19 subcategories), \keras (17 subcategories).
\tensor and \keras are popular libraries for deep learning, and above average interest in the majority
of the aspects of their functionality reflects their popularity.
\scikit is a popular ML library in Python. Though it is not 
used for deep learning, its use for regression, supervised and unsupervised 
learning, and recommendation related tasks are well known. This library 
provides abstract APIs that hides the details of ML.
In our study, the majority of questions about \scikit were about data 
preparation (26\%), modeling (25\%), and training (18\%).
In the middle, $>$30\% subcategories, we have \mllib (13 subcategories),  
\caffe (12 subcategories), \weka (11 subcategories), \theano (11 subcategories),
and \torch (9 subcategories). 
At the bottom, we have \mahout (6 subcategories) and \ho (8 subcategories).
We have previously observed that \mahout and \ho have questions observed under few
categories associated with initial stages. 
Combining with this observation suggests that such difficulties are higher
for \mahout and \ho compared to other libraries.

Next, we look at outliers.
For shape mismatch \keras is an outlier,
for model creation \mahout is an outlier,
for model selection \scikit is an outlier,
for output interpretation \ho and \keras are outliers,
for tuning strategy \ho is an outlier,
for tuning parameter selection \scikit is an outlier,
for model reuse \keras and \weka are outliers, and 
for bug \caffe and \scikit are outliers.

\findings{\scikit is an outlier in several categories suggesting that a deeper look into
	its API design might be necessary to improve usability of this important library.}{Blackbox abstraction for feature selection/extraction 
	APIs may not be suitable for ML libraries.} 

\scikit provides a lot of optional parameters to be selected in their APIs, whose values are hard to 
select yet affect accuracy. That could be the reason why its users have more 
difficulties in selecting parameters. 
\scikit also has an abnormally high percentage of questions about model selection,
which is surprisingly because it is one of the few libraries to provide abstract
model selection APIs, but the use of these APIs could be simplified.
This calls for research on designing better APIs for \scikit.

Next, we will look at the error/exception related questions.
\findings{Deep learning libraries \caffe, \ho, \keras, \tensor, \theano, \torch show more training time difficulties compared to other ML libraries}{arg2}
While this finding shouldn't be a surprise, it reinforces a well-established worry in both
the AI/ML and SE/PL communities that explaining why a deep learning model has worked or 
failed at training time or gives unexpectedly low performance remains a hard and open question.
We confirm that it is important to solve it to help developers make effective use of deep learning APIs.
To ensure that the dataset represents the usage of these libraries in the open source projects, we have calculated the number of occurrences of these libraries in Github open source projects. Table \ref{tbl:user} reports the number of occurrences of each library in GitHub. Furthermore, we performed the Kolmogorov Smirnov\cite{lilliefors1967kolmogorov} test among the distribution of the library usage population and our dataset population. We have found $p$-value of $0.675$ and $KS-statistics$ value as $0.3$, which suggest that both samples have been taken from a similar population.
 \begin{table}[t]
	\centering
\caption{Number of occurences utlizing the libraries in GitHub.}
\setlength{\tabcolsep}{4.8pt}
\begin{tabular}{ |l r | } 
 \hline
 \rowcolor{gray}
 \textcolor{white}{\bf Library}  & \textcolor{white}{\bf Occurences} \\ 
 \hline
 \hline
 
 \caffe & 1,46,121 \\ 
  \rowcolor{lightgray}
 \ho & 33,112  \\ 
\keras  & 7,55,427  \\ 

  \rowcolor{lightgray}
  \mahout & 2,793  \\ 
\mllib  & 90,042  \\ 
 
  \rowcolor{lightgray}
  \scikit & 2,69,672  \\ 
 
\tensor &  39,41,629  \\ 
 
  \rowcolor{lightgray}
  \theano& 2,28,960 \\ 
 
 \torch  & 1,21,583  \\ 
  \rowcolor{lightgray}
  \weka & 21,779 \\ 
 Overall  & 56,11,118 \\ 
 \hline
\end{tabular}
 \label{tbl:user}
\end{table}
\section{RQ4: Time consistency of difficulty}
\begin{figure}[ht]
	\includegraphics[width=\columnwidth]{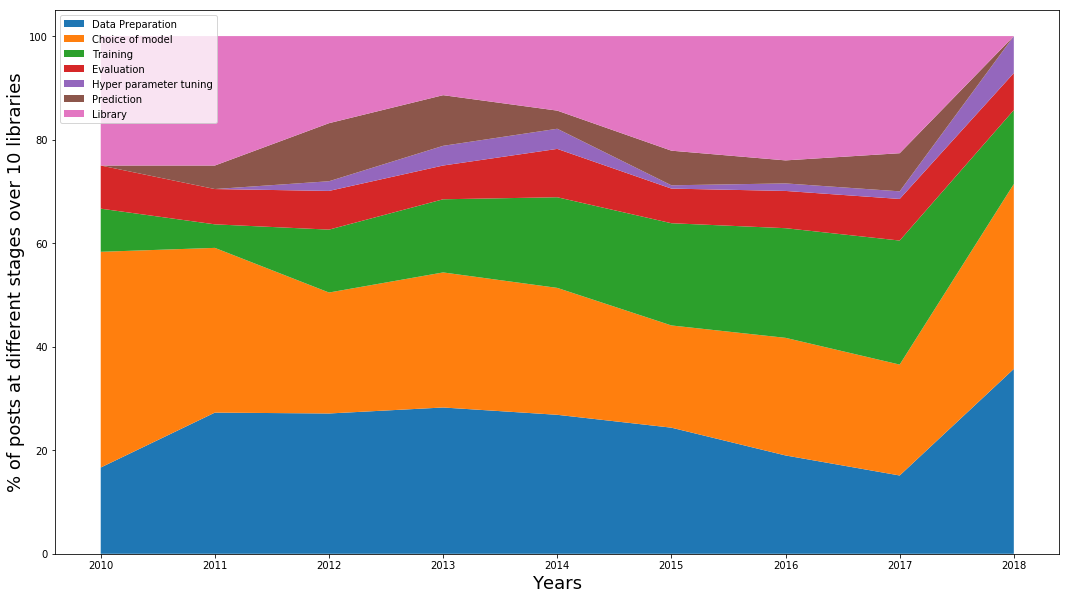}
	\caption{Difficulties over time, across different stages}
	\label{fig:cons}
\end{figure}
In this section we explore the answer to \textbf{RQ4} to understand whether the problems across different stages stayed consistent over time or are there problems that were prominent only for a certain period of time and then solved by the library developers. To study this question, we plot the percentage of posts across different stages of all the libraries from the year 2009 to March 2018. 

Our major observations from \fignref{fig:cons} are described below: 
\textbf{Model creation related problems are consistent over time.} Choice of model problems seem consistent over time indicating model creation problems are not being affected by the evolution of libraries. 
While these are fundamental problems for ML, 
deeper involvement of SE engineering researchers is needed to 
glean and disseminate lessons, patterns, and anti-patterns 
to help ML practice.

\textbf{Data preparation related problems slowly decrease after 2013 and show sharp increase after 2017.} Weka the library that has most difficulty in data preparation stage started losing popularity and new tensor representation of data gained popularity which explains the slow decline in the data preparation difficulty. 
The increase in data preparation since 2017 coincides with the increasing interest in deep learning,
and popularity of deep learning libraries that provide higher levels of abstraction.
Data from a varied set of sources are prepared for deep learning tasks. 

\textbf{Training related problems shows slow increase over time.} Due to the popularity of deep learning where training time errors occur more frequently the training related problems are slowly increasing.

\textbf{Evaluation problems are consistent over time.} Evaluation related problems have not been solved by the evolution of ML libraries over the last decade. 

	\section{Related Work}
\label{sec:related}
\sof is the widely used platform to study the software engineering
practice from the developer's perspective.
{\em However, existing work has not studied the usage of ML libraries using \sof.}
Meldrum {\em et. al.}~\cite{meldrum2017crowdsourced} studied 266 papers 
using \sof platforms to show the growing impact of \sof 
on software engineering research. 
Treude {\em et. al.}~\cite{treude2011programmers} did a manual
labeling of 385 questions to manually classify 385 questions into 10 different
categories (how-to, discrepancy, environment, error, decision help, conceptual,
review, non-functional, novice, and noise) to identify the question types. 
This study is useful to learn the general categories of questions asked by
developers. 
Kavaler {\em et. al.}~\cite{kavaler2013using} used \sof data to study 
the queries on APIs used by Android developers and showed the correlation 
between APIs used in producing Apps in the market and the questions on APIs 
asked by developers.
Linares-V{\'a}squez {\em et. al.}~\cite{linares2014api} studied the effect of
the changes in Android API on the developer community. They used the discussions
arising on \sof immediately after the API is changed and behavior of
the API is modified to study the impact of the change among the developers. 
\cite{berral2016quick} studied machine learning based algorithms, approaches, execution frameworks and presented a brief discussion of some libraries used in machine learning.
Barua {\em et. al.}~\cite{barua2014developers} studied the \sof posts 
and used  LDA topic modeling to extract topics to study the trend of different 
topics over time. 
Rebou{\c{c}}as {\em et. al.}~\cite{rebouccas2016empirical} studied the usage 
pattern of swift programming language among developers using \sof data. 
Schenk {\em et. al.}~\cite{schenk2013geo} studied the geographical distribution 
of usage and knowledge of different skills using \sof posts and users data. 
Stanley {\em et. al.}~\cite{stanley2013predicting} proposed a technique based 
on the Bayesian probabilistic model to predict the tags of a \sof post. 
McDonnel {\em et. al.}~\cite{mcdonnell2013empirical} presented a study of API 
stability using \sof data and as a test case they used Android Ecosystem.
Baltadzhieva {\em et. al.}~\cite{baltadzhieva2015predicting} proposed a
technique to predict the quality of a new \sof question. 
Joorabchi {\em et. al.}~\cite{joorabchi2016text} studied the challenges faced 
by computer science learners in different topics and subjects using the 
\sof data.
These works are orthogonal to ours.

	\section{Conclusion}
\label{sec:conclusion}

This work is motivated by the need to empirically understand the 
problems with usage of ML libraries. To understand the problems, we retrieved a significant 
dataset of Q\&A from \sof, classified these questions into categories
and subcategories and performed analysis from four viewpoints:
finding the most difficult ML stage, understanding the nature of problems, 
nature of libraries and studying whether the difficulties stayed consistent over time.
We found that model creation is the most difficult stage followed by data preparation.
We found that type mismatch, data cleaning and parameter selection are difficult
across all libraries.
We also found that initial stages are harder for \ho and \mahout,
and \scikit has proportionately higher problems in several subcategories.
Lastly, we observed that data preparation and training related problems are 
showing a sign of increase going forward.
These findings are a call to action for SE researchers 
as engineering of software with ML components is likely to 
be routine in the next decade.

	\ifCLASSOPTIONcompsoc
	\section*{Acknowledgments}
	\else
	\section*{Acknowledgment}
	\fi

We thank the anonymous reviewers for their valuable time and constructive feedbacks. We also thank PhD students Hamid Bagheri and Giang NGuyen to take immense pain and labeling the questions manually with great care and helping to resolve the difference through multiple meetings .This material is based upon work supported by the
National Science Foundation under Grant CCF-15-18897 and CNS-15-13263.  
Any opinions, findings, and conclusions or recommendations expressed 
in this material are those of the authors and do not necessarily reflect the 
views of the National Science Foundation.

	\ifCLASSOPTIONcaptionsoff
	\newpage
	\fi

	\bibliographystyle{IEEEtran}
	\bibliography{refs}

	\begin{IEEEbiography}    [{\includegraphics[width=1in,height=1.25in,clip,keepaspectratio]{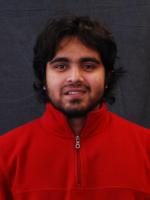}}]{Md Johirul Islam}
		is a doctoral candidate at
		Iowa State University. His research interests include machine learning program analysis, software techniques for machine learning, and programming languages. He has
		published works at FSE, Journal of Big Data Analytics in Transportation, and MSR.  
	\end{IEEEbiography}
	
	\begin{IEEEbiography}    [{\includegraphics[width=1in,height=1.25in,clip,keepaspectratio]{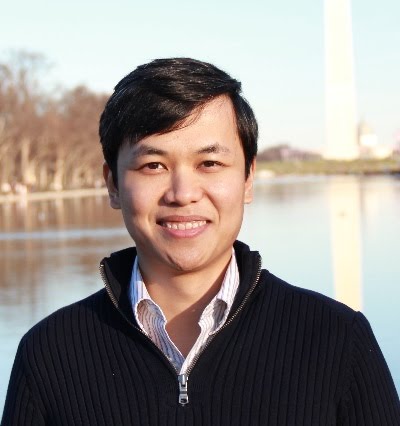}}]{Hoan Anh Nguyen}
		has done his post-doctoral at
		Iowa State University. He received his PhD from
		Iowa State University. His research interests include program analysis, software evolution and
		maintenance, and mining software repositories.
	\end{IEEEbiography}

\begin{IEEEbiography}    [{\includegraphics[width=1in,height=1.25in,clip,keepaspectratio]{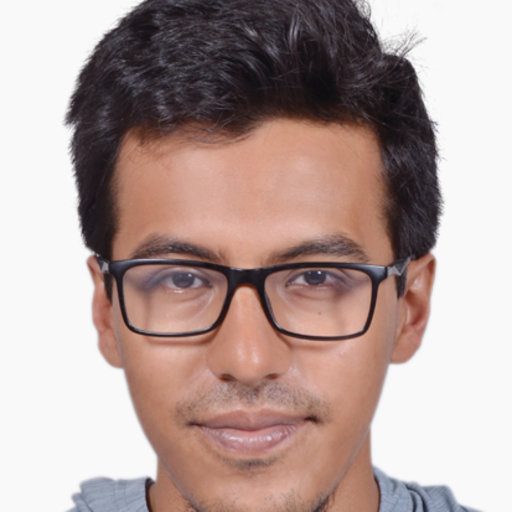}}]{Rangeet Pan}
	is a doctoral candidate at Iowa State University. His research interests include static analysis, machine learning, and software security. He has published work at FSE.
\end{IEEEbiography}

\begin{IEEEbiography}    [{\includegraphics[width=1in,height=1.25in,clip,keepaspectratio]{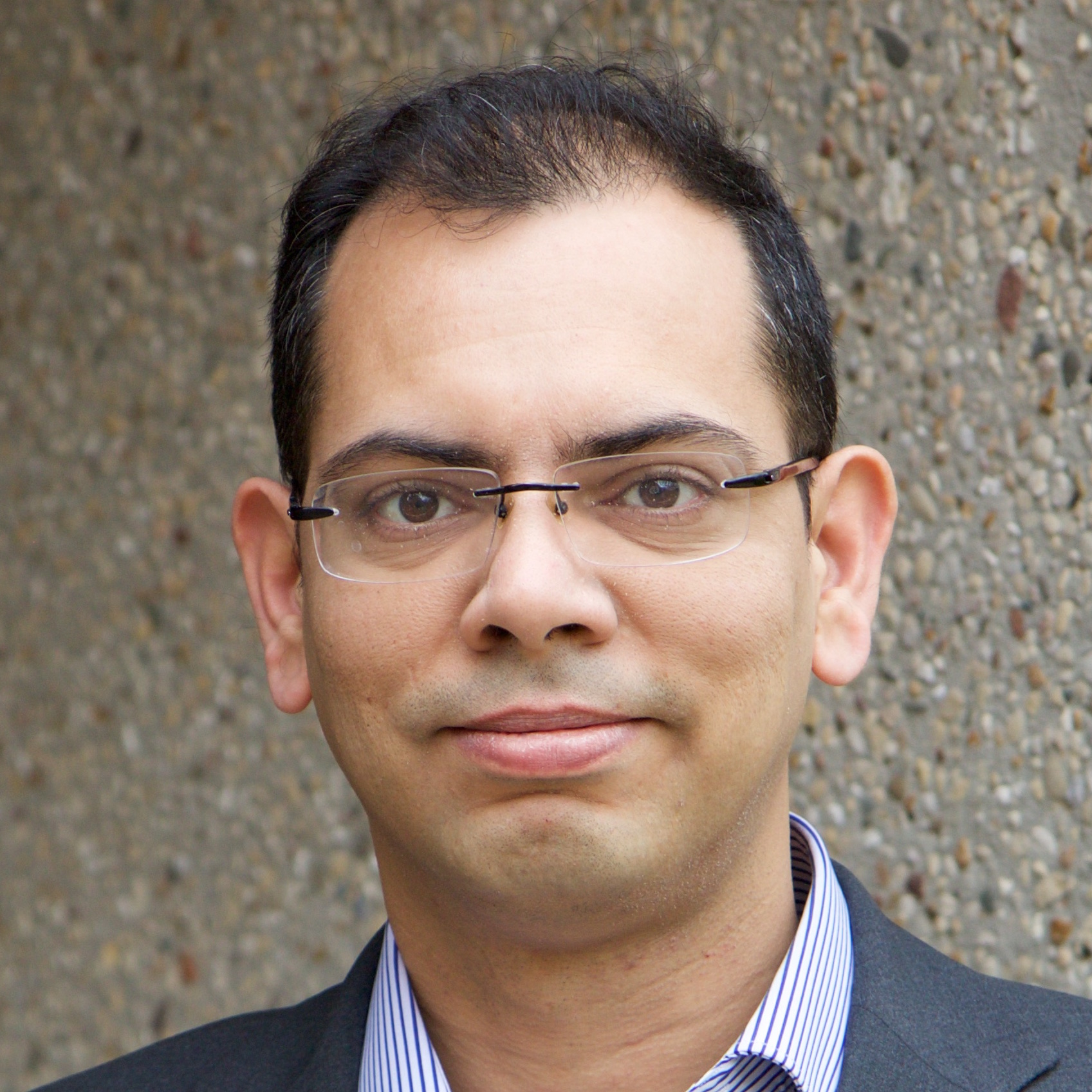}}]{Hridesh Rajan}
	is the Kingland Professor in the
	Computer Science Department at Iowa State
	University (ISU) where he has been since 2005.
	His research interests include programming languages, software engineering, and data science. He founded the Midwest Big Data Summer School to deliver broadly accessible data
	science curricula and serves as a Steering Committee member of the Midwest Big Data Hub
	(MBDH). He has been recognized by the US
	National Science Foundation (NSF) with a CAREER award in 2009 and by the college of LAS with an Early Achievement in Research Award in 2010, and a Big-12 Fellowship in 2012. He
	is a senior member of the ACM and of the IEEE.
\end{IEEEbiography}
	
	
	
	
	

\end{document}